\documentclass[11pt,reqno]{amsart}
\usepackage[english]{babel}
\usepackage{amsmath}
\usepackage{amssymb}
\usepackage{amsfonts}
\usepackage{graphicx}
\usepackage{xcolor}
\usepackage[colorlinks]{hyperref}

\UseRawInputEncoding

\begin{document}
\renewcommand{\refname}{References}
\renewcommand{\proofname}{Proof.}
\renewcommand{\figurename}{Fig.}

\title[Loss of hyperbolicity]{\large Numerical study of loss of hyperbolicity using a cold plasma model}


\author [Chizhonkov]{ E.V.~Chizhonkov $^1$}
\author[Rozanova]{ O.S. Rozanova$^1$}

\address[1]{Lomonosov Moscow State University}

\email{chizhonk@mech.math.msu.su, rozanova@mech.math.msu.su}



\begin{abstract}
We study a one-dimensional system of cold plasma equations taking into account electron-ion collisions in both relativistic and nonrelativistic cases. It is known that for a constant collision coefficient $\nu$, the solution to the Cauchy problem for such a system can lose smoothness. However, if the dependence of $\nu$ on the electron density $N$ is more than linear, then the solution remains globally smooth for any initial data. However, the appearance of the dependence $\nu(N)$ leads to a change in the type of the system, it loses hyperbolicity, which leads to computational problems.
In this paper, we propose a new implicit solution method in Euler variables that overcomes these difficulties. It can be used in both nonrelativistic and relativistic cases and is tested for the threshold  case of a linear dependence $\nu(N)=\nu_1+\nu_0 N$,
when smoothness can still be lost.
The computational experiments carried out are in full agreement with the available theoretical results.
\end{abstract}

\maketitle

\def\dfrac#1#2{\displaystyle{#1\over #2}}
\def\za{\alpha}
\def\zb{\beta}
\def\zp{\varphi}
\def\zt{\theta}
\def\zl{\lambda}
\def\bv{{\bf v}}
\def\zd{\delta}
\def\zg{\gamma}
\def\ze{\varepsilon}
\def\zr{\rho}
\def\zD{\Delta}

\section{Introduction}

We will consider nonlinear equations for cold plasma oscillations in the case where all physical quantities depend only on the coordinate $x$ and time $t$, and the vectors are oriented along the $x$ axis. In this case,
the system of hydrodynamic equations and Maxwell's equations for dimensionless values of the velocity $V=v_x/c$, momentum $P=p_x/(m_ec)$, and electron density $N=n_e/N_{0e}$, as well as the electric field $E=-eE_x/(m_ew_pc)$, taking into account collisions with a frequency $\nu_e$, can be represented as
\begin{eqnarray}
&&\dfrac{\partial  }{\partial \theta}  P +
V\dfrac{\partial }{\partial \rho} P= - E - \nu P,\qquad
\dfrac{\partial }{\partial \theta} E +
V\dfrac{\partial }{\partial \rho}  E = V, \label{u1}\\
&&N= 1 -
\dfrac{\partial   }{\partial \rho} E,\label{u1N}
\end{eqnarray}
	 Here $P, V, E, N $ are functions of $\zt$ and $\zr$, where $\zt = w_pt$, $\zr = k_p x$; in addition, $\nu = \nu_e/w_p$,
$w_p = \left(4 \pi e^2 N_{0e}/m_e\right)^{1/2}$ is the plasma frequency, $k_p = w_p/c$ , $N_{0e}$ is the electron density at equilibrium, $e, \,m_e$ are the charge and mass of the electron, and $c$ is the speed of light. The velocity of electrons is related to their momentum by the following relation
\begin{equation}\label{u11}	
V = P/ \zg, \quad \zg =\sqrt{1 + P^2}.
\end{equation}	
The quantity $\zg=\zg(\zr,\zt)$ is usually called the Lorentz factor.

The system of equations~(\ref{u1}), \eqref{u1N} is one of the simplest plasma models, often referred to as the equations
of <<cold>> plasma hydrodynamics;
it is well known and described in sufficient detail in textbooks and monographs (see, for example,~\cite{ABR78}, \cite{david72}, \cite{daw59}, \cite{GR75}, \cite{S71}, \cite{SR12}).

Electron-ion collisions are described by the term $\nu P$ in the momentum equation.
This effect can be interpreted as a frictional force between particles; in the nonrelativistic case (see, for example, ~\cite{ABR78}),
they use a formula
$$
- \nu_{\za\zb} \left( \bv_{\za} - \bv_{\zb} \right),
$$
where $\nu_{\za\zb}$ is the effective collision frequency of charged particles of type $\za$ with particles of type $\zb$, when $\za \neq \zb$.
For stationary ions ($\bv_{\zb}=0$), the formula is simplified. A detailed description of the formulas for plasma transport coefficients is presented in ~\cite{b-20}.
If $\nu$ is constant, then the equations \eqref{u1} can be considered independently of \eqref{u1N}. The system \eqref{u1}
is non-strictly hyperbolic according to the standard classification (e.g., ~\cite{RYa}). Namely, the matrix of derivatives has a single eigenvalue, but there is a complete set of eigenvectors forming a basis for the space. The properties of such systems are well studied, and it is known that the solution of the Cauchy problem for such systems with smooth initial data can develop a singularity in finite time, associated with either the solution itself or its derivatives becoming infinite \cite{Daf16}. It is easy to see that the solution of the system \eqref{u1} itself is bounded, but the derivatives can become infinite. Moreover, as can be seen from \eqref{u1N}, a singularity forms at the component of the electron density.
In plasma physics, this effect is usually called  {\it breaking}  of oscillations.
As shown in \cite{ZM},
the singularity that arises when the electron density tends to infinity in the Eulerian description of the medium's motion is
equivalent to the intersection of electron trajectories in its Lagrangian description.

 For $\nu=0$, in \cite{RChDAN20} (see \cite{RChZAMP21} for detailed proofs), a criterion for singularity formation in terms of initial data for the nonrelativistic case was obtained. For the relativistic case, in \cite{RChZAMP26}, it was shown that for general initial data, a singularity always forms.
 However, it is always possible to find a class of initial data for which the existence of a smooth solution continues for an arbitrarily large predetermined time. For $\nu={\rm const}>0$, a criterion for singularity formation in the nonrelativistic case was obtained in \cite{RChD20}; in
\cite{RChZAMP21fr}, this problem was considered for the relativistic case.
The results of these studies included the construction and validation of high-precision algorithms
for the numerical simulation of nonlinear oscillations of cold plasma using Lagrangian variables~\cite{ChDR2012}.
Approximate methods in second- and third-order Eulerian variables are also known~\cite{Ch20},~\cite{Ch23GVM};
however, modeling the breaking effect using them is complicated by the presence of large gradients of the solution
near the singular density values.

We note that the solutions of the Vlasov kinetic equation, from which the system \eqref{u1}, \eqref{u1N} is obtained in a certain limiting case, do not exhibit the breaking effect~\cite{iord61}. Therefore, various attempts have been made to regularize the cold plasma equations; for example, it has been proposed
to include electron-ion collisions in the model along with viscosity (see~\cite{infeld},\cite{verma}).
Of course, this led to a smoothing of the solutions due to a change in the type of equations being solved (from hyperbolic to parabolic),
but at the same time, the physical model of the process was qualitatively changed.
n ~\cite{BS2017}, for the non-relativistic case, a regularization was proposed using the electron-ion collision coefficient, which was considered a linear function of the electron density, which is quite consistent
with the physics of the phenomenon~\cite{b-20}. Using numerical experiments on solutions linear in the spatial coordinate (they are also called affine or axial~\cite{Ch_book}), it was shown in~\cite{BS2017} that the linear dependence of the collision frequency on the density does indeed prevent the occurrence of the overturning effect, in contrast to cases where the coefficient $\nu$ is a positive constant or equal to zero.
However, it was subsequently shown in~\cite{RozNHYP} that the effect of complete regularization for the case of a linear dependence $\nu(N)$ exists only for affine solutions, and in the general case, to completely suppress the breaking, one must require a dependence $\nu(N)$ stronger than linear (for example, $\nu=\ze N^\gamma$, $\gamma>1$). These results were obtained for the nonrelativistic case, but it can be shown that they remain valid in the relativistic case as well.

Thus, the most physically reasonable case
\begin{equation}\label{funcnu}
\nu (N) = \nu_0 N + \nu_1.
\end{equation}
where $\nu_0, \nu_1$ are some non-negative constants, is a threshold case, and it is natural to attempt a numerical study of the influence of the coefficients on the properties of the solution.

However, it should be noted that the linear dependence of the collision coefficient on density qualitatively changes the mathematical properties of the cold plasma model, namely, the system of equations \eqref{u1} ceases to be hyperbolic.
Indeed, if we substitute the value of $N$ from \eqref{u1N} into \eqref{u1}, it is easy to see that the matrix of derivatives becomes a Jordan block. This matrix still has two coinciding eigenvalues, but they correspond to only one eigenvector. That is, the system \eqref{u1} becomes structurally the same as the famous <<pressureless>> gas dynamics system (e.g., \cite{ZM}), which can develop a strong singularity in the solution component (i.e., in $V,E$, not in $N$).
This automatically calls into question the applicability of traditionally used difference methods for numerical solutions~\cite{KPS12}. Of course, the smallness of $\nu_0$ allows us to hope for similar properties of the solutions of the original hyperbolic and perturbed systems, but this requires rigorous proof. In any case (whether the solutions are close or significantly different), approximate methods for solving new nonlinear formulations that are perturbations of hyperbolic systems are of particular interest. The primary motivation for this research was the numerical solution of nonstandard equations.

We study the Cauchy problem for~(\ref{u1}), (\ref{u11}) under initial conditions corres\-pon\-ding to
plasma oscillations localized in space near the the axis $\zr = 0$. We assume that the initial velocity and momentum of the electron are zero, that is,
\begin{equation}\label{u12}	
V(\zr,\zt=0) = 0, \quad P(\zr,\zt=0) = 0,
\end{equation}	
and the oscillations are excited at the initial moment of time by an electric field of the following type~\cite{Ch_book}:
\begin{equation}\label{gauss}
E(\rho,\zt=0) = \za \, \rho \, \exp\left\{-2
\dfrac{\rho^2}{\rho_*^2}\right\},\quad \za = \left(\dfrac{a_*}{\rho_*}\right)^2,
\end{equation}
where the parameters $\zr_*$ and $a_*$ characterize the scale of the localization region and the maximum value $E_{\max} = a_*^2/(\zr_*2\sqrt{\rm e}) \ \approx 0.3a_*^2/\zr_*$ of the electric field~(\ref{gauss}), respectively.
In accordance with the form of the electric field~(\ref{gauss}), the electrons are initially shifted from the axis $\zr = 0$ in different directions, which subsequently leads to their oscillations relative to this axis. The form of the function~(\ref{gauss}) is chosen based on the fact that such oscillations can be excited in a rarefied plasma $(w_l \gg w_p$) by a laser pulse with frequency $w_l$ when it is focused into a line (this can be achieved using a cylindrical lens~\cite{Shep13}).

We obtain a nonrelativistic approximation of the system \eqref{u1}. Since, under the assumption that the electron velocity $V$ is small, \eqref{u11} implies the represen\-tation
$
P= V + \dfrac{V^3}{2} + O(V^5),\quad V\to 0,
$
then, up to cubically small terms, we can assume that $P=V$. This assumption allows us to write \eqref{u1}, \eqref{u1N} in the form
\begin{equation}\label{u2}
\dfrac{\partial V }{\partial \theta}  +
V\,\dfrac{\partial V}{\partial \rho}+ E + \nu V = 0,\quad
\dfrac{\partial E }{\partial \theta} +
V\,\dfrac{\partial E}{\partial \rho} - V
=0, \quad N = 1 - \dfrac{\partial E}{\partial \rho}.
\end{equation}
In this case, the initial conditions \eqref{u12} remain unchanged.

We  consider the dimensionless collision frequency $\nu$ in the form \eqref{funcnu}.
In the literature, the case $\nu_1>0$, $\nu_0 = 0 $, is most often considered; here, however, it is useful for comparison with the case $\nu_1=0$, $\nu_0 > 0 $,
which is the main focus of our attention.

The paper is organized as follows.
Section 1 proposes a numerical algorithm similar to the implicit MacCormack scheme for solving relativistic equations. This algorithm generalizes the well-known explicit MacCormack scheme and is easily adapted to the nonrelativistic case.
Section 2 is devoted to numerical experiments. It is shown that a linear dependence of the collision coefficient on the electron density leads to a stronger damping of the oscillation amplitude compared to the absence of this dependence in both the relativistic and nonrelativistic cases.
It is further established that the influence of electron-ion collisions manifests itself in a slowdown of the breaking process of multi-period oscillations, even to the point of its complete elimination.
The Conclu\-sions systematize the results of the conducted research.


\section{Implicit MacCormack-type scheme for relativistic equations}

Taking into account the explicit expression for the electron density $N(\rho,\theta)$ in \eqref{funcnu},
we  reduce  system~(\ref{u1}) to a vector form convenient for the case under consideration
\begin{equation}
\label{systvec}
\dfrac{\partial {\bf U}}{\partial \zt} + A(V,\zg)\,\dfrac{\partial {\bf U}}{\partial \zr} =
 {\bf  S}({\bf U}),
\end{equation}
where the matrix
\begin{equation}
\label{matrixA}
A(V,\zg) = V
\begin{pmatrix}
1 & - \nu_0 \zg\\
0 & 1
\end{pmatrix}
\end{equation}
has the form of a second-order Jordan block
for $\nu_0 \neq 0$,
${\bf U}=(P,E)^T$, ${\bf S}=(-E -(\nu_0 + \nu_1)P,V)^T$ are vector functions considered
in the half-plane $\{(\zr,\zt)\,:\, \zt \ge 0,\, \zr \in {\mathbb R}\}$.
The matrix $A(V,\zg)$ has two identical real eigenvalues, but only one eigenvector, i.e.
\eqref{systvec} is not of the hyperbolic type, so traditional
numerical methods oriented toward hyperbolic conservation laws~\cite{Daf16} should be applied to \eqref{systvec}
with great caution.

We define a discretization of the independent variables using constant parameters $\tau$ and $h$ such that
$$
\zt^n = n\,\tau,\; n \ge 0,\quad \zr_{i} = ih, \; i = 0, \pm 1, \pm 2, \dots
$$
and we denote the dependent variable ${\bf U}(\zr,\zt)$ at the grid node $(\zr_i,\zt^n)$ as ${\bf U}_i^n$.

Let us introduce useful notations for the forward difference operators $D^{+}$ and backward difference operators $D^{-}$, for which the argument can be either vector or scalar:
$$
D^{+}{\bf F}_{i}={\bf F}_{i+1} - {\bf F}_{i}, \quad D^{-}{\bf F}_{i}={\bf F}_{i} - {\bf F}_{i-1},
$$
and write an implicit MacCormack-type scheme for system~(\ref{systvec}), assuming that $P_i^n, \, E_i^n$ are known:

\noindent
1) We define a predictor with the result ${\bf U}_i^p$ using the following algorithm:

First, we calculate $\zg_i^n = \sqrt{1+\left(P_i^n\right)^2}, \; V_i^n = P_i^n/\zg_i^n$, then for

\noindent $V^n_{i+1/2} = \left( V^n_{i+1} + V^n_{i} \right)/2$ we define the matrix
$$
C^n_{i+1/2} = A\left(|V^n_{i+1/2}|, \zg_i^n\right)
$$
and sequentially calculate
\begin{equation}
\label{predMC}
\begin{array}{c}
\zD {\bf U}_i^n = - \dfrac{\tau}{h} \,A\left(V^n_{i+1/2}, \zg_i^n\right)\, D^{+}{\bf U}_{i}^n
+\tau {\bf S}_{i}^n, \; \;  
\left(I - \zl \dfrac{\tau}{h}\,C^n_{i+1/2}\, D^{+}\right) \zd {\bf U}_i^p = \zD {\bf U}_i^n,\vspace{0.5 ex}\\
{\bf U}_i^p = {\bf U}_i^n + \zd {\bf U}_i^p.
\end{array}
\end{equation}

\noindent 2) We define a corrector with the result ${\bf U}_i^c$ using the following algorithm:

We first calculate $\zg_i^p = \sqrt{1+\left(P_i^p\right)^2}, \; V_i^p = P_i^p/\zg_i^p$, then for

\noindent $V^p_{i-1/2} = \left( V^p_{i-1} + V^p_{i} \right)/2$ we define the matrix
$$
C^p_{i-1/2} = A\left(|V^p_{i-1/2}|,\zg_i^p\right)
$$ and sequentially calculate
\begin{equation}
\label{corrMC}
\begin{array}{c}
\zD {\bf U}_i^p = - \dfrac{\tau}{h} \,A\left(V^p_{i-1/2}, \zg_i^p\right)\, D^{-}{\bf U}_{i}^p
  + \tau {\bf S}_{i}^p, \; \; 
\left(I + \zl \dfrac{\tau}{h}\,C^p_{i-1/2}\, D^{-}\right) \zd {\bf U}_i^c = \zD {\bf U}_i^p, \vspace{0.5 ex}\\
{\bf U}_i^c = {\bf U}_i^n + \zd {\bf U}_i^c.
\end{array}
\end{equation}
In formulas~(\ref{predMC}),~(\ref{corrMC})
the superscript $p$ (or $c$) denotes the predictor (or corrector) step, or $n$ denotes the time layer $\zt^n$, and $\zl$ denotes a constant parameter of the circuit, which will be defined below.

The final formulas, generating the solution at the next time layer with the number $(n+1)$, are as follows:
 $$
 {\bf U}_i^{n+1} = \dfrac{{\bf U}_i^p  + {\bf U}_i^c}{2}, \quad
N^{n+1}_i = 1 - \dfrac{E_{i+1}^{n+1}-E_{i-1}^{n+1}}{2h}.
$$
The scheme can be applied to solving the system~(\ref{u2}) (the nonrelativistic case) if we set $\zg_i^n = \zg_i^p \equiv 1$ in formulas~(\ref{predMC}),~(\ref{corrMC}).

Note that in the case of $\nu_0=0$ (the diagonal matrix $A$ in~(\ref{matrixA})), the scheme has proven itself to be excellent
for both the relativistic equations~(\ref{u1}) and the simplified nonrelativistic analog~(\ref{u2}) (see~\cite{Ch24VMPr}, \cite{Ch24IzvV}).
These same papers also describe the properties of the scheme using the example of a scalar transport equation, including
approximation, stability, and implementation features.

Note that for $\zl = 0$, the proposed implicit scheme~(\ref{predMC}),~(\ref{corrMC}) transforms into the well-known explicit
MacCormack scheme~\cite{mcc2003}, adapted for modeling plasma oscillations~\cite{Ch20}.
Furthermore, if the matrix $A$ has the shape of a Jordan block, then the transition operator in the difference scheme, responsible for
its stability, will have a similar form. For a sufficiently large number of time steps, the norm of the transition operator
will exceed unity, which will lead to an increase in the computational error for any choice of grid parameters.
In other words, the choice of an implicit scheme (for example, for $\zl=1$) in this case is dictated by the need for stable calculations
over long time intervals for a system of equations that is not hyperbolic, but close to it
(due to $0 < \nu_0 \ll 1$).


Let us illustrate this with a simple example. Consider a natural implementation of the well-known <<upwind>> scheme
in the presence of a second-order Jordan block:
\begin{equation}
\label{primer1}
\begin{array}{c}
\dfrac{u_m^{n+1}-u_m^{n}}{\tau} + a\,\dfrac{u_m^{n}-u_{m-1}^{n}}{h} + b\,\dfrac{v_m^{n+1}-v_{m-1}^{n+1}}{h} = 0, \vspace{0.5 ex}\\
\dfrac{v_m^{n+1}-v_m^{n}}{\tau} + a\,\dfrac{v_m^{n}-v_{m-1}^{n}}{h} = 0,
\end{array}
\end{equation}
where the constants $a > 0, \, b \ge 0, \, a \gg b$ are given, and $\tau$ and $h$, as above, determine the discretization parameters.

To analyze the stability of the  scheme
~(\ref{primer1}), we use the spectral stability criterion~\cite{BKCh}.
Let
$$
u_m^n = c_1 q^n \exp^{{\rm \bf i} m \zp}, \quad v_m^n = c_2 q^n \exp^{{\rm \bf i} m \zp},
$$
where ${\rm \bf i}$ is the imaginary unit,
and we write the second relation~(\ref{primer1}) as
\begin{equation*}
\label{primer2}
v_m^{n+1} = q\, v_m^{n}, \quad q = 1 - \dfrac{\tau \, a}{h} \left( 1 - \exp^{- \,{\rm \bf i} \zp} \right).
\end{equation*}
Then the first relation~(\ref{primer1}) can be rewritten as
\begin{equation*}
\label{primer3}
u_m^{n+1} = q\, u_m^{n} + q\, y\, v_m^{n}, \quad y = -\,\dfrac{\tau \, b}{h} \left( 1 - \exp^{- \,{\rm \bf i}  \zp} \right),
\end{equation*}
which allows us to define the transition operator $G$ in the
scheme
~(\ref{primer1}):
\begin{equation*}
\label{primer4}
\begin{pmatrix}
u_m^{n+1}\\
v_m^{n+1}
\end{pmatrix} = G
\begin{pmatrix}
u_m^{n}\\
v_m^{n}
\end{pmatrix}, \quad
G = q
\begin{pmatrix}
1 & y\\
0 & 1
\end{pmatrix}.
\end{equation*}
Consider $\|G\|_2^2$, the square of the spectral norm~\cite{BKCh} of the matrix $G$,
equal to the maximum eigenvalue of the matrix $G^*G$; recall that this norm is subject to the usual Euclidean vector norm.
Simple calculations lead to the expression
\begin{equation*}
\label{primer5}
\|G\|_2^2 = |q|^2 \left[ 1 + |y|^2/2 + |y| \sqrt{1 + |y|^2/4} \right].
\end{equation*}
It is well known (see, for example,~\cite{BKCh}) that for $b=0$, which yields $y=0$, the scheme~(\ref{primer1}) is stable for $\tau \,a /h \le 1$, that is
$$
\|G\|_2^2 = |q|^2 \le 1 \quad \forall \zp \in [0, 2\pi).
$$
In the considered case of a Jordan block (for $ b > 0 $) we have
$$
|y|= 2\, \dfrac{\tau \, b}{h} \left| \sin (\zp/2) \right|,
$$
which for $\zp = \pi, \, \tau \, a = h $ yields
$$
\|G\|_2^2 > 1 + 2\, \dfrac{\tau \, b}{h} = 1 + 2\, \dfrac{b}{a},
$$
that is, instability of the  scheme~(\ref{primer1}).

It can also be shown that even a more general choice of $\tau$ and $h$ does not guarantee the formal stability of this scheme. Indeed,
consider the case of small positive angles $\zp$, then the leading terms of the asymptotics (linear with respect to $\zp$) will have the form
$$
|q|^2 \approx 1, \quad
|y| \approx  \dfrac{\tau \, b}{h} \zp .
$$
This leads to the expression
$$
\|G\|_2^2 \approx 1 + \dfrac{\tau \, b}{h} \zp = 1 + c\, \dfrac{\tau}{h}
$$
with a constant $c$ independent of the discretization parameters.
The spectral stability criterion is not satisfied here, so for small ratio  $b/a$  and over short time intervals,
the scheme~(\ref{primer1}) can, of course, be used, but with caution; for a large number of time steps, an implicit (more stable!) difference scheme is recommended.

\section{Numerical experiments}

For numerical simulation of plasma oscillations, the computational domain must be bounded;
we define it as a segment $[-d,d]$, at the endpoints of which artificial boundary conditions must be specified.
Section 3.6 of the book~\cite{Ch_book} is devoted to a discussion of their construction; here, we restrict ourselves to
<<cutting off>> the infinite domain using homogeneous boundary conditions of the first kind:
$$
 P(\pm d,\theta) = V(\pm d,\theta) =  E(\pm d,\theta) = 0.
$$
Of course, the parameter $d$ should be chosen sufficiently large.
Due to the exponential decay of the function $E_0(\rho)$,
it is sufficient to set $d = 4.5 \rho_*$. In this case, we have
$\exp^2 \{- d^2/\rho_*^2\} \approx 2.5768 \cdot 10^{-18}$. This means that, when calculating
with double precision, the magnitude of the jump in the initial function $E_0$ at the points $\rho = \pm d$ is commensurate with machine precision,
i.e., with the usual rounding error. In other words, when numerically modeling
the oscillations, the effect of truncating the initial conditions will be completely unnoticeable, which
fully corresponds to the concept of an <<artificial bounda\-ry>>.

\subsection{Non-relativistic calculations}

\begin{center}
\begin{figure}[h!]
\includegraphics[scale=0.7]{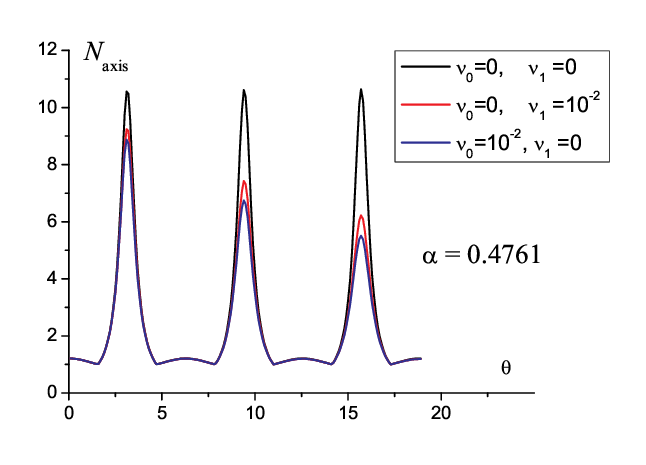}
\caption{Dynamics of density in nonrelativistic oscillations for $\za=0.4761$: the breaking effect is absent for different collision coefficients}
\label{pic1}
\end{figure}
\end{center}

We fix the parameters $a_*=0.414,$ $\rho_*=0.6$. Their choice is significant primarily for the range of electron density:
the electron concentration in the center of the region can be many times greater than the equilibrium (background) value of unity.
 These para\-me\-ters lead to low-intensity oscillations when collisions are absent: the maximum density is only approximately 10 times greater than the background value.
Given the exponential nature of the initial perturbation in~(\ref{gauss}), we choose sufficiently small discretization parameters:
$\tau = h = 1/4000$. This allows us to keep the error for the least smooth electron density function $N(\zr,\zt)$ within
a tenth of a percent where it is possible to compare the numerical and analytical solutions (see, for example,~\cite{Ch24VMPr}).
Recall that the MacCormack scheme (both explicit and implicit) has, under the stability conditions,
an asymptotic order of accuracy of $O(\tau^2 + h^2)$, so the indicated choice of grid steps is quite sufficient for adequately reproducing the solution (see.~\cite{Ch20}).
Fig.\ref{pic1} illustrates the collision models under consideration for the chosen $a_*$ and $\rho_*$; they correspond to
the value $\za=0.4761$. Recall that in the absence of collisions, the chosen $\za$ satisfies the condition
of the global existence of a solution~\cite{RChDAN20},~\cite{RChZAMP21}, the black graph in Fig. \ref{pic1} corresponds to
a $2\pi-$periodic perturbation of the electron density.
As shown in~\cite{RChD20}, if a global solution without collisions exists, then a global solution
with any positive constant collision coefficient also exists. The red graph in Fig.\ref{pic1} corresponds to the case $\nu_0=0, \, \nu_1=10^{-2}$. The time-damped oscillations of the electron density are completely consistent with the available theoretical results.
Finally, the blue color in Fig.\ref{pic1} denotes the case of a linear dependence of the collision coefficient on density
($\nu_0=10^{-2}, \, \nu_1=0$). For general initial conditions, this case does not yet have a complete theoretical justification~\cite{RozNHYP},
although it is intuitively clear that a variable collision coefficient should enhance the damping of the oscillation amplitude compared to a constant coefficient when the solution is sufficiently smooth, i.e., when it exists globally without taking collisions into account.
Next, with the same value of $\rho_*=0.6$, we increase the value of $a_*$ in~(\ref{gauss}) to obtain $\za=0.51$.

\begin{center}
\begin{figure}[h!]
\includegraphics[scale=0.7]{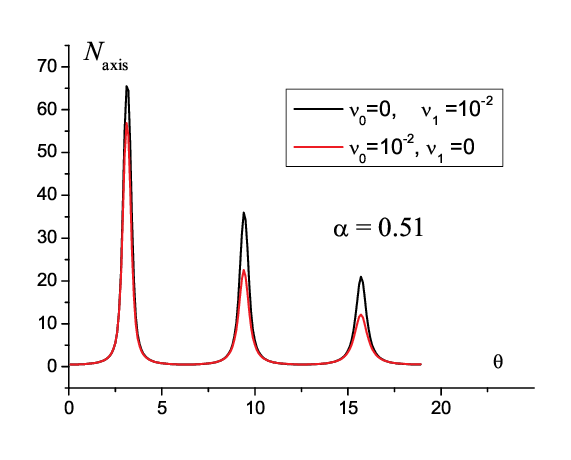}
\caption{Dynamics of density in nonrelativistic oscillations for $\za=0.51$: the breaking effect is absent only for non-zero collision coefficients}
\label{pic2}
\end{figure}
\end{center}

This value leads to the breaking of oscillations without taking into account collisions, but their amplitude will be limited when
the collision coefficient is constant ($\nu_0=0, \, \nu_1=10^{-2}$). Fig.\ref{pic2} illustrates this situation.
Note a characteristic difference from Fig.\ref{pic1}, a significant increase in the oscillation amplitude (approximately sevenfold).
However, by virtue of the existence theorem~\cite{RChD20}, we observe a monotonic attenuation of the oscillation amplitude over time
at a constant collision coefficient (black graph), in full agreement with theory.
More interesting is the observation that, similar to the case with a globally smooth solution,
oscillations with a variable collision coefficient ($\nu_0=10^{-2}, \, \nu_1=0$) also decay faster than those with a constant coefficient. In this case, this fact is experimental and does not yet have a full theoretical basis.

Based on numerical experiments conducted for nonrelativistic equations~(\ref{u2}), we can formulate the hypothesis
that the use of the variable collision coefficient $\nu = \nu_0 N$, $\nu_0>0$, leads to an expansion of the set of initial data leading
to the existence of a globally smooth solution, compared to the case of a constant coefficient $\nu = \nu_1$.
Additionally, we note that for the given set of initial data~(\ref{u12}),~(\ref{gauss})
in the case of the collision coefficient $\nu = \nu_0 N$, the breaking effect never occurs, i.e., the electron
density does not have a singularity for any $\nu_0 > 0$. It can be assumed that this is related to the specific type of solution, as was the case for axial (affine) solutions~\cite{RozNHYP}.

\subsection{Relativistic calculations}
To demonstrate the proposed algorithm (\ref{predMC}), (\ref{corrMC}) for solving the relativistic problem \eqref{u1}, we present calculations of the breaking effect, which can be observed after several oscillation periods.
We fix the parameters $a_*=3.105$, $\rho_*=4.5$ in order to satisfy the condition of a local
in-time existence of a solution for the system \eqref{u1} at $\nu=0$~\cite{RChDAN20},~\cite{RChZAMP21}, and also to maintain continuity with the results of the numerical and asymptotic analysis from~\cite{FrCh}.

Calculations for system \eqref{systvec}  were performed with $\tau = h = 1/2000$ until
the electron density function became infinite. To ensure accuracy, calculations were also performed with grid parameters
half as small as the main (working) ones. The results of the calculations in Eulerian variables are completely consistent with those obtained using high-precision methods based on Lag\-ran\-gian variables~\cite{ChDR2012} for all the experiments presented below
both without collisions and using a constant coefficient $\nu_1$.

\begin{center}
\begin{figure}[h!]
\includegraphics[scale=0.7]{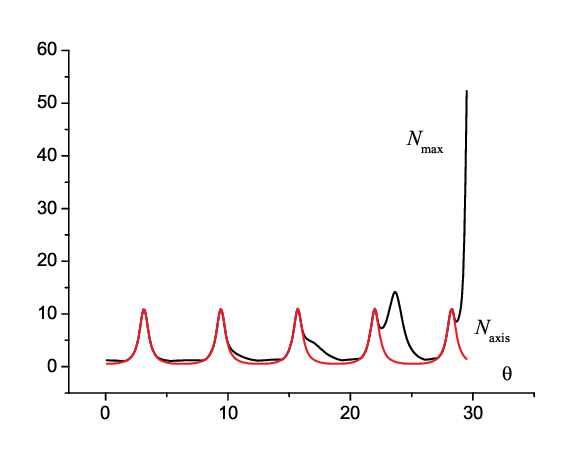}
\caption{Dynamics of density in relativistic  collisionless plasma: maximum over the region (black) and at the origin (red)}
\label{pic3}
\end{figure}
\end{center}

\begin{center}
\begin{figure}[h!]
\includegraphics[scale=0.7]{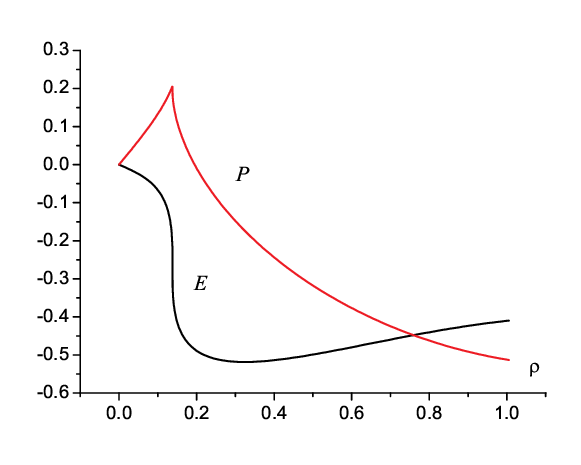}
\caption{Momentum and electric field at the moment of breaking in a relativistic collisionless plasma}
\label{pic4}
\end{figure}
\end{center}

For a constant collision coefficient, in~\cite{FrCh} asymptotic methods were used to establish the dependence of the breaking time
on  $ \nu_1 \zt^{(0)}_{wb}$, where $\zt^{(0)}_{wb}$ is the breaking time of oscillations without taking collisions into account, but with the same parameters $a_*$ and $\rho_*$.
For comparison,  Fig.\ref{pic3} shows the dependence of the maximum of density  on time in a collisionless plasma,
i.e., for $\nu_0=0, \, \nu_1=0$. It follows from this that, for the given calculation parameters, the maximum of density  off the oscillation axis is formed in the fourth oscillation period, which already in the next period at $\zt^{(0)}_{wb} \approx 29.5$ has an infinite value.

Fig. \ref{pic4} shows the spatial distributions of the momentum $P$ and the electric field $E$ in a collisionless plasma at the moment of breaking  $\theta \approx 29.5$, when the absolute maximum density outside the origin becomes unbounded. Note that, due to the structure of equations \eqref{u1}
their solutions $P$ and $E$ remain odd functions of the coordinates if the
initial data have this property (see~\cite{Ch24IzvV}). The initial data
\eqref{gauss} are precisely such, therefore, the functions are shown only on the positive semi-axis.
We emphasize that the breaking has the character of a <<gradient catastrophe>>, i.e., the functions $P$ and $E$ themselves remain bounded~\cite{RYa}. A discussion of the nature of the discontinuity can be found in \cite{RChZAMP26}.

\begin{center}
\begin{figure}[h!]
\includegraphics[scale=0.7]{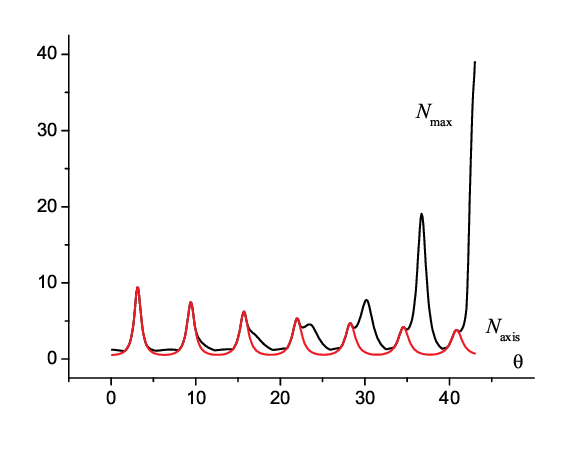}
\caption{Dynamics of density in relativistic oscillations taking into account collisions ($\nu_0=0, \, \nu_1 \zt^{(0)}_{wv}=0.3$): maximum over the region (black) and
at the origin (red).}
\label{pic5}
\end{figure}
\end{center}

When electron collisions are taken into account, the time for the appearance of a density singularity increases; we present the results for the case $\nu_0=0$.
For example, for $\nu_1 \zt^{(0)}_{wb} = 0.3$, the off-axis maximum (see Fig. \ref{pic5} ) reaches infinity only in the fourth period after its formation, and not in the second period, as was the case in the collisionless case. Calculations show that, for given initial parameters, a singularity of density arises only in the case of relatively rare collisions, when the inequality
$\nu_1 \zt^{(0)}_{wb} \le 0.422$ is satisfied.
When the equality $ \nu_1 \zt^{(0)}_{wb} = 0.422$ holds, the time it takes for the singularity to appear is $ \zt_{wb} \approx 75.22$, which is approximately 2.5 times longer than the breaking time in a collisionless plasma.
When the condition $ \nu_1 \zt^{(0)}_{wb} > 0.422$ is satisfied, the density singularity no longer occurs.
In this case, the off-axis maximum initially increases after its formation, but then decreases due to the strong damping of the oscillations~\cite{ChDR2012},~\cite{FrCh}.

Let us estimate the nonlinear oscillation breaking time for some typical plasma parameters~\cite{FrCh}. If plasma oscillations are excited by a gaussian electric field with parameters $a_*=3.105,$ $\rho_*=4.5$, then the breaking time in a collisionless plasma is $ \zt^{(0)}_{wb} \approx 29.5$, as follows from calculations. Hence, in accordance with the numerical result for the threshold value of the dimensionless collision frequency $ \nu_1 \zt^{(0)}_{wb} \le 0.422$, it follows that the breaking effect occurs when the inequality  $\nu_1 \le 1.43 \cdot 10^{-2}$ is satisfied. In a fully ionized plasma, the dimensionless frequency of electron-ion collisions is determined by the formula~\cite{SR12}
 \begin{equation}\label{fornu}
\nu = Z \dfrac{\sqrt{8}}{3}\eta^{3/2} \ln \Lambda,
\end{equation}
where $Z$ is the ion charge number, $\ln \Lambda$ is the Coulomb logarithm, and $\eta$ is the ratio of the electron interaction energy
$e^2 N_{0e}^{1/3}$ to the electron kinetic energy $T_e$,
 \begin{equation}\label{foreta}
\eta =  \dfrac{e^2 N_{0e}^{1/3}}{T_e}.
\end{equation}

\begin{center}
\begin{figure}[h!]
\includegraphics[scale=0.7]{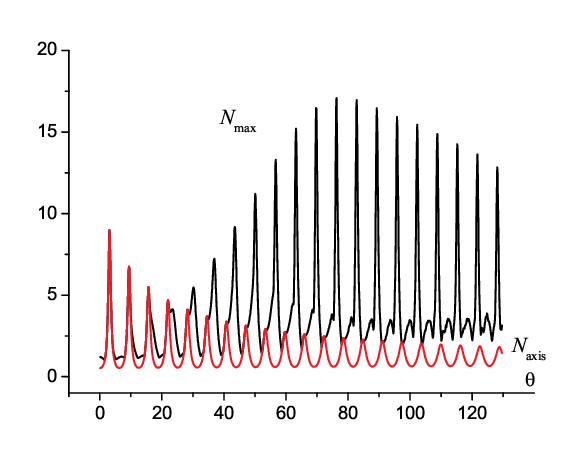}
\caption{Density dynamics in relativistic oscillations taking into account collisions ($\nu_0 \zt^{(0)}_{wv}=0.3, \, \nu_1=0$): maximum over the region (black) and
at the origin (red)}
\label{pic6}
\end{figure}
\end{center}

Let a laser pulse with a wavelength of $\lambda=1.24$ $\rm \mu m$ (frequency $\omega_l \approx 1.5 \cdot 10^{15} {\rm s}^{-1}$), duration $\tau \approx 36$ $\rm fs$ and dimensionless electric field amplitude equal to $2.5$
propagate in a rarefied, fully ionized plasma with an ion charge $Z=5$, electron density $N_{0e}=10^{18}{\rm cm}^{-3}$, and temperature $T_e = 50$ $\rm eV$.
If a laser pulse is focused by a cylindrical lens into a line with a transverse dimension of $L_x \approx 24$ $\rm \mu m$, then in the wake wave behind the pulse, mainly flat one-dimensional oscillations of electrons with parameters
$a_*=3.1,$ $\rho_*=4.5$
are excited,
which are close to the values used above in the calculations.

Note that when a laser pulse with moderately relativistic intensity $a_0 \approx 1-3$ propagates in a rarefied plasma, the condition for optimal excitation of plasma waves~\cite{GrUr2017} is approximately preserved, and the amplitude of plasma oscillations is related to the laser field by the same relation $a_*^2 \approx 1.52 \,a_0^2$ as in the nonrelativistic limit. For the given plasma parameters
from formulas \eqref{fornu}, \eqref{foreta} we find that the dimensionless collision frequency $\nu_1$
is less than the threshold value $1.43 \cdot 10^{-2}$ . Therefore, in this case, the breaking effect occurs, and the breaking time
is approximately equal to $ \zt_{wb} \approx 43$, since the parameter $ \nu_1 \zt^{(0)}_{wb} \approx 0.3$. If we consider the propagation of a laser pulse with the above parameters in a plasma with the same density, but with a temperature of $T_e = 20$ eV, then calculations using the formulas \eqref{fornu}, \eqref{foreta} yield the following value for the dimensionless collision frequency: $\nu_1 \approx 1.8 \cdot 10^{-2}$
(parameter $ \nu_1 \zt^{(0)}_{wb} \approx 0.52$), which exceeds the threshold value. Therefore, in the case of such a choice of parameters, the breaking of plasma oscillations in the wake wave of the laser pulse does not occur due to strong attenuation~\cite{FrCh}.

For the case of a variable collision coefficient ($\nu_0 > 0, \nu_1=0$), the breaking effect is not observed under weaker constraints.
Figure \ref{pic6} shows the dynamics of density for the parameters $\nu_0 \zt^{(0)}_{wb} = 0.3, \nu_1=0$. It is easy to see that
the absolute values of maxima
of the electron density do not exceed $20$, although they are near this value for approximately ten consecutive periods. Meanwhile, the axial density values decrease monotonically.

Based on the numerical experiments conducted for relativistic equations~(\ref{u1}), we can formulate the hypothesis
that the use of a variable collision coefficient $\nu = \nu_0 N$
makes it possible to significantly increase the oscillation breaking time compared to the case $\nu = \nu_1$, $\nu_0 = 0$.
In other words, the regularization of cold plasma hydrodynamic equations proposed in paper~\cite{BS2017} is very useful
and interesting from a mathematical point of view, but this cannot completely prevent the breaking of oscillations.

In conclusion, we emphasize that long-term calculations were performed only using the implicit scheme, while calculations over relatively
short time intervals ($5-7$ periods) were performed using both the implicit and implicit schemes. Moreover, the calculation results
for identical grid parameters diffe\-red little from each other.


\section*{Conclusion}
In this paper, a new implicit second-order solution method in Euler variables is constructed for one-dimensional plane cold plasma equations, taking into account electron-ion collisions. This method can be used both with and without taking relativistic effects into account.

The main feature of the proposed method is its applicability when the original hydrodynamic model
of cold plasma loses its hyperbolic type and the  matrix of derivatives does not have a complete set of eigenvectors.

The computational experiments performed are fully consistent with existing theoretical results obtained by both analytical methods (nonrelativistic case) and asymptotic methods (relativistic case).

Furthermore, a numerical analysis of the new model (with a linear dependence of the collision coefficient on the electron density) revealed that in the relativistic case, the well-known effect of breaking multi-period plasma oscillations is observed, and it was confirmed that it has the character of a gradient catastrophe.
It was also shown that taking electron-ion collisions into account leads to a smoothing of the solution when relativity effects are neglected, and also
slows down the process of breaking relativistic oscillations to the point of its complete elimination (for a sufficiently large $\nu_0$).

These results can be used in numerical modeling of laser-plasma interactions, as well as for developing a theory of numerical methods for solving singularly perturbed hyperbolic systems.

\end{document}